\documentclass[twocolumn,showpacs,superscriptaddress,preprintnumbers,amsmath,amssymb]{revtex4}
\usepackage{graphicx}% Include figure files
\usepackage{dcolumn}% Align table columns on decimal point
\usepackage{bm}% bold math

\begin{document}

\title{Phonon dispersion curves and thermodynamic properties of $\alpha$-Pu$_{2}$O$_{3}$}

\author{Yong Lu}
\affiliation{LCP, Institute of Applied Physics and Computational
Mathematics, Beijing 100088, China}

\author{Yu Yang}
\affiliation{LCP, Institute of Applied Physics and Computational
Mathematics, Beijing 100088, China}

\author{Fawei Zheng}
\affiliation{LCP, Institute of Applied Physics and Computational
Mathematics, Beijing 100088, China}

\author{Ping Zhang}
\thanks{Author to whom correspondence should be
addressed. E-mail: zhang\_ping@iapcm.ac.cn} \affiliation{LCP,
Institute of Applied Physics and Computational Mathematics, Beijing
100088, China} \affiliation{Beijing Computational Science Research Center, Beijing 100089, China}

\date{\today}% It is always \today, today,

\begin{abstract}

A recent inelastic x-ray scattering study [Manley et al., Phys. Rev.
B {\bf 85}, 132301 (2012)] reveals that the phonon dispersion curves
of PuO$_2$ is considerably consistent with our previous density
functional +$U$ results [Zhang et al., Phys. Rev. B {\bf 82}, 144110
(2010)]. Here in the present work, using the same computational
methods, we further obtain the phonon dispersion curves for
$\alpha$-Pu$_2$O$_3$. We find that the Pu-O bonding is weaker in
$\alpha$-Pu$_{2}$O$_{3}$ than in fluorite PuO$_{2}$, and
subsequently a frequency gap appears between the vibrations of
oxygen and plutonium atoms. Based on the phonon dispersion curves
and Helmholtz free energies of PuO$_2$ and $\alpha$-Pu$_{2}$O$_{3}$,
we systematically calculate the reaction energies for the
transformations between Pu, PuO$_2$, and $\alpha$-Pu$_2$O$_3$. It is
revealed that the thermodynamic equilibrium of the system is
dependent on temperature as well as on the chemical environment.
High temperature and insufficient oxygen environment are in favor of
the formation of $\alpha$-Pu$_2$O$_3$.

\end{abstract}


\maketitle

\section{INTRODUCTION}

Plutonium-based materials attract much interest not only owing to
their industrial, military, and environmental importance, but also
for their basic theoretical prospects. Due to the complex character
of Pu 5$f$ electrons, which locates in the boundary of localized and
delocalized among the actinide metals, Pu occurs in six phases under
different temperatures and pressures \cite{Savrasov,Moore}. When
metallic Pu is exposed to dry air at room temperature, the
protective plutonium dioxide PuO$_{2}$ layer is formed, which can
further reduce to another thin layer of plutonium sesquioxide
$\alpha$-Pu$_{2}$O$_{3}$ at the oxide-metal interface
\cite{Haschke}. Since plutonium oxides can be applied to the nuclear
stockpile and storage of surplus plutonium, revealing the physical
properties of different plutonium oxides and the thermodynamic
equilibrium criteria between them become necessary and important. In
particular, the temperature dependence of the free energy of
different plutonium oxides are key factors for their thermodynamic
equilibrium.

The phase diagram of stoichiometric plutonium-oxygen system shows
the presence of plutonium dioxide PuO$_{2}$ and sesquioxide
Pu$_{2}$O$_{3}$. PuO$_{2}$ shares the crystal structure of fluorite
type, isomorphous with $An$O$_{2}$ ($An$=U, Th and Np). The actinide
sesquioxide crystallizes into three different crystal structures,
the hexagonal La$_{2}$O$_{3}$ structure, the monoclinic
Sm$_{2}$O$_{3}$ structure, and the cubic Mn$_{2}$O$_{3}$ structure,
respectively \cite{Petit10}. Pu$_{2}$O$_{3}$ has been synthesized
only in the Mn$_{2}$O$_{3}$ ($\alpha$-Pu$_{2}$O$_{3}$) and
La$_{2}$O$_{3}$ structures ($\beta$-Pu$_{2}$O$_{3}$) \cite{Petit10}.
The $\alpha$-Pu$_{2}$O$_{3}$ is stable only below 300 $^{\circ}C$.
Structurally, the $\alpha$-Pu$_{2}$O$_{3}$ unit cell can be obtained
by ordered removal of 25 percent of the oxide ions from PuO$_{2}$
2$\times$2$\times$2 supercell. Due to the structural similarity with
PuO$_{2}$, $\alpha$-Pu$_{2}$O$_{3}$ is more likely to be the
reduction product of PuO$_2$. In fact, a mixture of these two oxides
can be prepared by partial reduction of PuO$_{2}$ at high
temperature and then cooling to room temperature \cite{IAEA}.

Because of its structural complexity (64 atoms in each unit cell),
$\alpha$-Pu$_2$O$_3$ has seldom been theoretically studied.
Nevertheless, it is key to understand the thermodynamic equilibrium
of the plutonium surface under different oxidizing environments.
Therefore in the present work, we systematically study the physical
and thermodynamic properties of $\alpha$-Pu$_2$O$_3$, the
temperature dependent reaction energy between $\alpha$-Pu$_2$O$_3$
and PuO$_2$ is also discussed.

Apart from the structural complexity of Pu$_2$O$_3$, modeling of the
electron localization/delocalization of plutonium 5$f$ electrons is
also a complex task. Conventional density functional theory (DFT)
schemes that apply the local density approximation (LDA) or the
generalized gradient approximation (GGA) underestimate the strong
on-site Coulomb repulsion of the plutonium 5\emph{f} electrons and
consequently fail to capture the correlation-driven localization.
However, the DFT+$U$ method proposed by Dudarev \emph{et al.}
\cite{Dudarev1, Dudarev2} is an effective way to deal with the
strong on-site Coulomb repulsion of Pu, U and Np 5\emph{f}
electrons. In our previous studies, we have verified the validity
and reliability of this method more than once
\cite{Zhang,Shi,Wang,Lu,Sun}, and the obtained phonon dispersion
curves of PuO$_2$ are very comparable with latest experiments
\cite{Manley}.

\section{Calculation methods}

The density functional theory (DFT) calculations are carried out
using the Vienna \textit{ab initio} simulations package (VASP)
\cite{G.Kresse1,G.Kresse2} with the projected-augmented-wave (PAW)
potential method \cite{PAW}. The plane-wave basis set is limited
with an energy cutoff of 500 eV. The exchange and correlation
effects are described by generalized gradient approximation (GGA) in
the Perdew-Burke-Ernzerhof (PBE) form \cite{PBE}. The computational
cells for $\alpha$-Pu$_{2}$O$_{3}$ and PuO$_{2}$ contain 64 and 12
atoms respectively. And integration over their Brillouin Zones are
done on 5$\times$5$\times$5 and 15$\times$15$\times$15 $k$-ponit
meshes generated by using the Monkhorst-Pack method
\cite{Monkhorst}, respectively. These $k$-point meshes are carefully
tested and proven to be sufficient for energy convergence of less
than 1.0$\times$10$^{-4}$ eV per atom.

The strong on-site Coulomb repulsion among the localized Pu 5$f$
electrons is described by the GGA+$U$ formalisms formulated by
Dudarev \emph{et al.},\cite{Dudarev1,Dudarev2}. As concluded in some
previous studies, although the pure GGA or LDA fail to depict the
electronic structure, especially the insulating nature and the
occupied-state character of PuO$_{2}$ and $\alpha$-Pu$_{2}$O$_{3}$
\cite{Sun,Zhang,Jomard,Shi}, the LDA/GGA+$U$ approaches will capture
the Mott insulating properties of the strongly correlated Pu 5$f$
electrons in plutonium oxides adequately. Since only the difference
between the spherically averaged screened Coulomb energy $U$ and the
exchange energy $J$ is significant for the total LDA/GGA energy
functional, we labeled them as one single effective parameter $U$
for simplicity. The value of exchange energy $J$ is set to be a
constant 0.75 eV, which is the same as in our previous study of
plutonium oxides,\cite{Sun,Zhang,Shi} while $U$ can suitably be
chosen within a range of values ($\sim$1-4 eV), which are in the
ballpark of commonly accepted values for actinide elements.

\section{Results and discussions}

\begin{figure}
\includegraphics[width=0.5\textwidth]{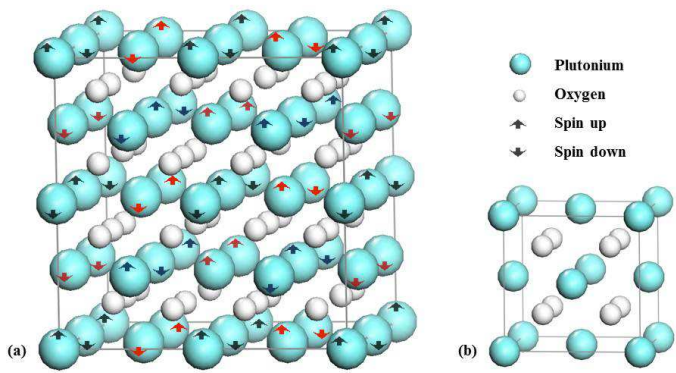}
\caption{(a) Atomic structure and magnetic arrangement of
$\alpha$-Pu$_{2}$O$_{3}$ within one unit cell. (b) Atomic structure
of the fluorite PuO$_2$. \label{fig:fig1}}
\end{figure}

\subsection{Ground-state properties of $\alpha$-Pu$_{2}$O$_{3}$ and PuO$_2$}

For GGA/LDA, Coulomb repulsion of the 5$f$ electrons in Pu describes
PuO$_{2}$ as incorrect ferromagnetic (FM) conductor instead of
antiferromagnetic (AFM) Mott insulator reported by experiment
\cite{McNeilly}. Our previous studies show that the DFT+$U$ approach
can accurately describe the atomic and electronic structures of
plutonium dioxide \cite{Zhang,Sun}. The AFM phase begins to be
energetically preferred at Hubbard $U$$\sim$1.5 eV, consequently,
AFM phase becomes more stable than FM phase above $U$=2 eV. As shown
in Fig. 1, we take use of the collinear AFM order for
$\alpha$-Pu$_{2}$O$_{3}$ unit cell and 2$\times$2$\times$2 PuO$_2$
supercell as proposed by Regulski \emph{et al.} \cite{Regulski},
since we have confirmed that this spin order is energetically
favored \cite{Shi}. We reported in Table I our calculated results of
lattice parameter $a_{0}$, bulk modulus $B_{0}$, pressure derivative
of the bulk modulus $B_{0}'$, spin moments $\mu_{mag}$, and energy
band gap $E_{g}$ for AFM $\alpha$-Pu$_{2}$O$_{3}$ and PuO$_{2}$ at 0
GPa within GGA and GGA+$U$ ($U$=3 eV). For comparison, experimental
values in Refs. \cite{IAEA,Zachariasen,Haschke,Idiri} are also
listed. All the values of $a_{0}$, $B_{0}$, and $B_{0}'$ are
obtained by fitting the third-order Brich-Murnaghan equation of
state \cite{Brich}. The lattice parameter $a_{0}$ of
$\alpha$-Pu$_{2}$O$_{3}$ is 11.169 \AA$~$at $U$=3 eV, in accordance
with the measured value range 11.03-11.07
\AA$~$\cite{Zachariasen,IAEA}. Comparing the volumes of PuO$_{2}$
2$\times$2$\times$2 supercell and $\alpha$-Pu$_{2}$O$_{3}$ unit
cell, there is about 7\% expansion when the reduction process
accomplishes. The bulk modulus $B_{0}$ of $\alpha$-Pu$_{2}$O$_{3}$
are calculated to be 120 GPa, which is much lower than the 192 GPa
value of PuO$_{2}$, indicating that the resistance to external
pressure of $\alpha$-Pu$_{2}$O$_{3}$ is weaker than PuO$_{2}$.

An accurate description of the electronic structure for materials is
of great importance. The photoemission experiments show that the
PuO$_{2}$ is Mott insulator \cite{Butterfield,Gouder}. In our
previous studies, the insulating nature and the occupied-state
character of plutonium dioxide, which the pure GGA fail to depict,
can be prominently improved by turning the effective Hubbard $U$
parameter in a reasonable range. When we turned on $U$ parameter,
the amplitude of insulating gap increases with enhancing $U$ for
$\alpha$-Pu$_{2}$O$_{3}$. Similar to PuO$_{2}$, the FM conductor to
AFM insulator transition of $\alpha$-Pu$_{2}$O$_{3}$ also occurs,
and at $U$=3 eV the insulating gap is 1.0 eV. The calculated
amplitude of local spin moment is 5.02 $\mu_{B}$ per Pu atom for AFM
$\alpha$-Pu$_{2}$O$_{3}$ within GGA+$U$ scheme, somewhat larger than
4.14 $\mu_{B}$ of PuO$_{2}$.

\begin{table}[ptb]
\caption{Calculated lattice parameters $a_{0}$, bulk modulus
$B_{0}$, pressure derivative of the bulk modulus $B_{0}'$, spin
moments $\mu_{mag}$, and energy band gap $E_{g}$ for AFM
$\alpha$-Pu$_{2}$O$_{3}$ and PuO$_{2}$ in the GGA and GGA+$U$ ($U$=3
eV) approaches at 0 GPa. For comparison, experimental values are
also listed.} \label{point}
\begin{ruledtabular}
\begin{tabular}{ccccccccccccc}
    &Method&$a_{0}$&$B_{0}$&$B_{0}'$&$\mu_{mag}$&$E_{g}$\\
    & &({\AA})&(GPa)&&($\mu_{B}$)&(eV)\\
\hline
$\alpha$-Pu$_{2}$O$_{3}$ & GGA      &10.921&123&4.25&4.95&0.0\\
                         & GGA+$U$  &11.169&120&3.53&5.02&1.0\\
                         & Expt.    &11.03$-$11.07$^{a,b}$&&& \\
\hline
PuO$_2$ &GGA     &5.396&185&4.29&4.09&0.0\\
        &GGA+$U$ &5.457&192&4.50&4.14&1.2\\
        &Expt.   &5.398$^{c}$&178$^{d}$\\
\end{tabular}
\begin{flushleft}
$^{a}$ Reference \onlinecite{IAEA}\\
$^{b}$ Reference \onlinecite{Zachariasen}\\
$^{c}$ Reference \onlinecite{Haschke}\\
$^{d}$ Reference \onlinecite{Idiri}
\end{flushleft}
\end{ruledtabular}
\end{table}

\begin{figure}
\includegraphics[width=0.4\textwidth]{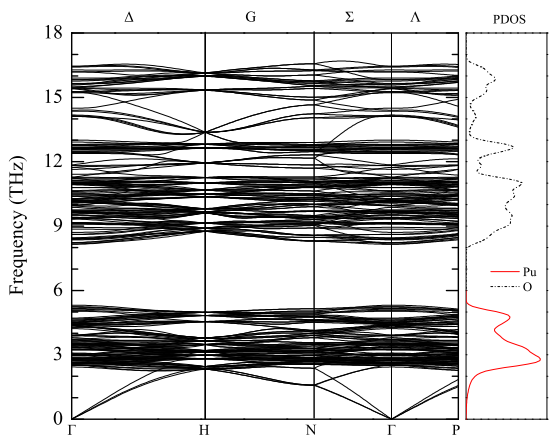}
\caption{Phonon dispersion curves (left) and phonon density of
states (Phonon DOS) (right) of $\alpha$-Pu$_{2}$O$_{3}$.
\label{fig:fig2}}
\end{figure}

\begin{figure}
\includegraphics[width=0.4\textwidth]{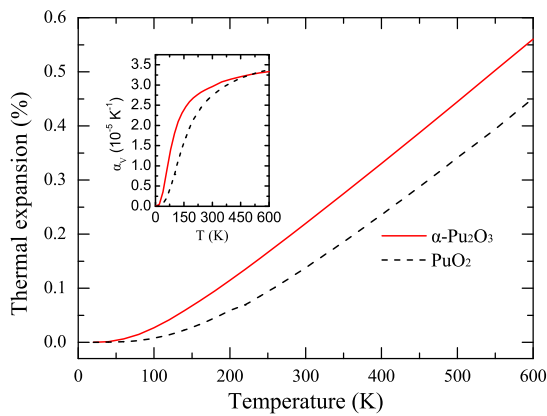}
\caption{Temperature dependence of the linear thermal expansion for
$\alpha$-Pu$_{2}$O$_{3}$ and PuO$_{2}$. The inset is the volume
thermal expansion coefficient as a function of
temperature.\label{fig:fig3}}
\end{figure}

\begin{figure}
\includegraphics[width=0.4\textwidth]{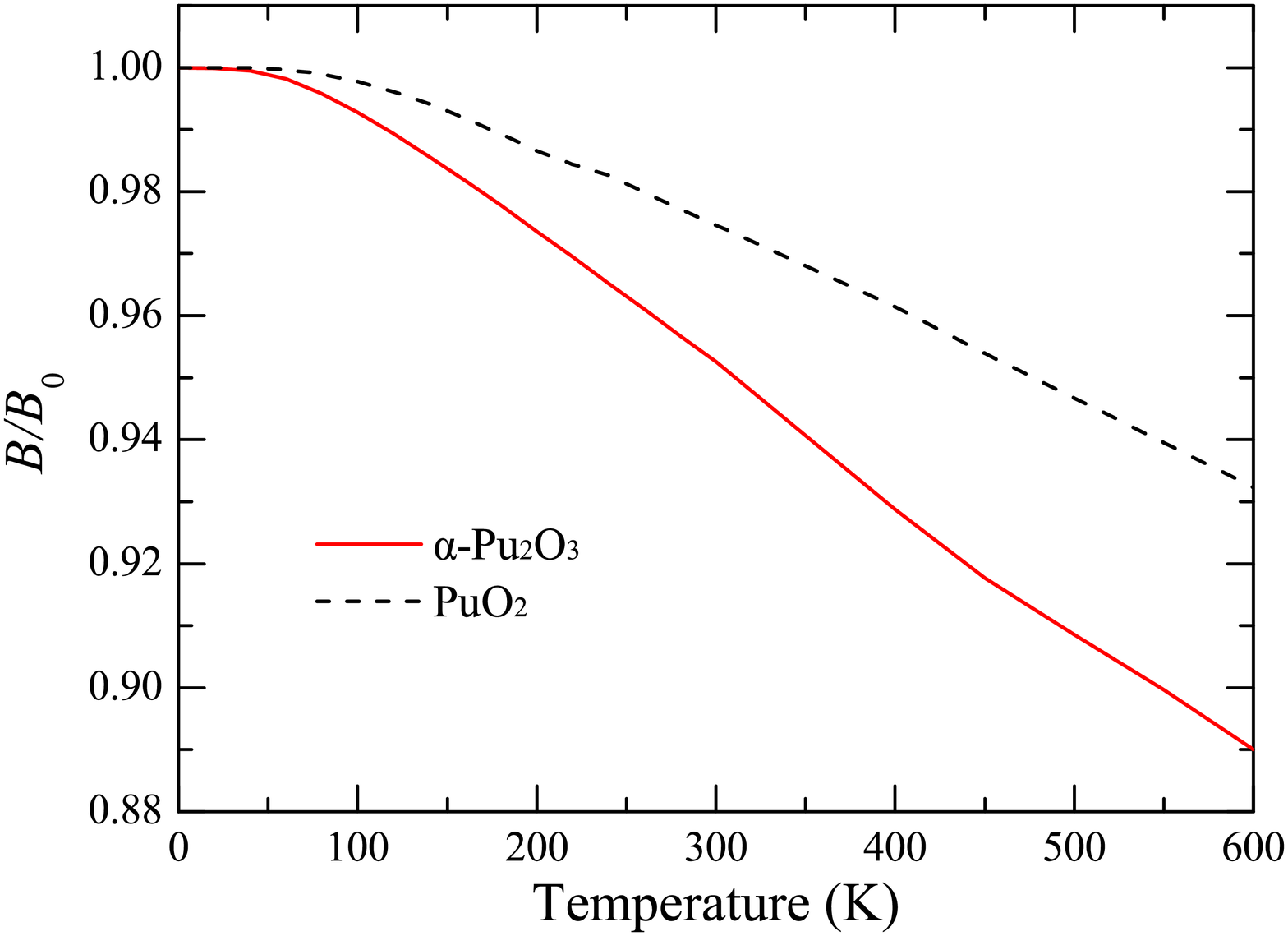}
\caption{Temperature dependence of the bulk modulus $B$ for
$\alpha$-Pu$_{2}$O$_{3}$. The inset is the ratio of $B/B_{0}$ for
$\alpha$-Pu$_{2}$O$_{3}$ and PuO$_{2}$.\label{fig:fig4}}
\end{figure}

\begin{figure}
\includegraphics[width=0.4\textwidth]{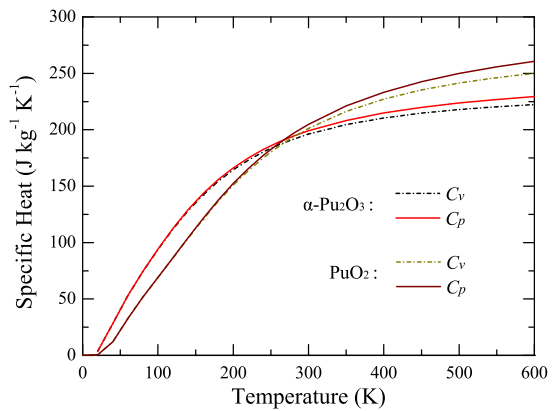}
\caption{Heat capacities of $\alpha$-Pu$_{2}$O$_{3}$ and PuO$_{2}$
at constant volume ($C_{V}$) and constant pressure ($C_{p}$).
\label{fig:fig5}}
\end{figure}

\subsection{Phonon dispersions, thermal expansion, and heat
capacity of $\alpha$-Pu$_{2}$O$_{3}$}

The phonon dispersion curves of a material is closely related to its
structure stability and thermodynamic properties like thermal
expansion, Helmholtz free energy, and heat capacity. In our present
study, the phonon frequencies are carried out using the direct
method. For calculating the Hessian matrix of
$\alpha$-Pu$_{2}$O$_{3}$ we adopted its unit cell with the size of
over 11 \AA along the $x, y, z$ directions, which is large enough to
isolate direct interactions of neighboring cells. And a
3$\times$3$\times$3 Monkhorst-Pack \emph{k}-point mesh is used for
integration over the Brillouin Zone while calculating the forces.
The calculated phonon dispersion curves of $\alpha$-Pu$_{2}$O$_{3}$
along the high-symmetry lines of its Brillouin Zone are shown in
Fig. 2, together with the atomic phonon density of state
(phonon-DOS). As presented, there is an evident frequency gap
between the oxygen and plutonium vibrations. The vibrations of the
plutonium and oxygen atoms dominate the low (0-5.2 THz) and high
frequency (8-16.5 THz) parts respectively. The obvious frequency gap
between oxygen and plutonium vibrations makes $\alpha$-Pu$_2$O$_3$
different from PuO$_2$, as there is a small overlap between oxygen
and plutonium vibration modes in PuO$_2$ \cite{Zhang,Manley}.

From Fig. 2 we see that the frequency value of the phonon dispersion
curves for $\alpha$-Pu$_2$O$_3$ can only be as large as 16.63 THz,
which is much smaller than the maximum value of 19.19 THz for
PuO$_2$. Since the high-frequency part corresponds to oxygen
vibrations, this result indicates that the binding of oxygen atoms
around plutonium atoms is stronger in PuO$_2$ than in
$\alpha$-Pu$_2$O$_3$. In addition, as mentioned above, the
$\alpha$-Pu$_{2}$O$_{3}$ unit cell can be obtained by orderly
removing 25 percent of the oxide atoms from a 2$\times$2$\times$2
PuO$_{2}$ supercell. The removal of oxygen atoms leads to a 7\%
volume expansion of $\alpha$-Pu$_2$O$_3$, and then weakens the bond
strength of Pu-O bondings. Overall, the all-positive phonon
dispersion curves confirm the thermodynamic stability of
$\alpha$-Pu$_{2}$O$_{3}$ at its ground state.

Based on the obtained phonon dispersion curvs, we can further
calculate the thermodynamic properties of $\alpha$-Pu$_{2}$O$_{3}$,
which begins by calculating its Helmholtz free energy $F(V, T)$.
Normally the of Helmholtz free energy of an insulator can be divided
into two parts as
\begin{equation}
F(V,T)=E(V)+F_{vib}(V,T), \label{eq1}
\end{equation}
where $E(V)$ is the ground-state electronic energy, and
$F_{vib}(V,T)$ stands for the phonon free energy at a given
temperature $T$. Within quasi-hamonic approximation (QHA), the
$F_{vib}(V,T)$ term can be evaluated by
\begin{equation}
F_{vib}(V,T)=k_{B}T\sum_{j,\mathbf{q}}\ln\left[ 2\sinh\left(
\frac{\hbar\omega_{j}(\mathbf{q},V)}{2k_{B}T}\right)  \right],
\label{eq2}
\end{equation}
where $\omega_{j}(\mathbf{q},V)$ is the phonon frequency of the
$j$th phonon mode with wave vector $\mathbf{q}$ at fixed cell volume
$V$, and $k_{B}$ is the Boltzmann constant. The total specific heat
of the crystal is the sum of all phonon modes over the BZ,
\begin{equation}
C_{V}(T)=\sum_{j,\mathbf{q}}c_{V,j}(\mathbf{q},T). \label{eq3}
\end{equation}
$c_{V,j}(\mathbf{q},T)$ is the mode contribution to the specific
heat defined as,
\begin{equation}
c_{V,j}(\mathbf{q},T)=k_{B}\sum_{j,\mathbf{q}}\left(\frac{\hbar\omega_{j}(\mathbf{q},V)}{2k_{B}T}\right)^{2}\frac{1}{\sinh^{2}[\hbar\omega_{j}(\mathbf{q},V)/2k_{B}T]}.\label{eq4}
\end{equation}

From the obtained phonon dispersion curves and Eqs. (1) and (2), the
Helmholtz free energy of $\alpha$-Pu$_{2}$O$_{3}$ is calculated at
different temperatures and different lattice constants. By
determining the minimum energy lattice constants at different
temperatures, the lattice expansion curve is obtained and shown in
Fig. 3. The corresponding result of PuO$_{2}$ is also shown in Fig.
3 for comparisons. One can see that the rate of lattice expansion is
larger for $\alpha$-Pu$_{2}$O$_{3}$ than for PuO$_{2}$, indicating
that the volume of $\alpha$-Pu$_{2}$O$_{3}$ is more sensitive to
temperature. This character can also be seen from the thermal
expansion coefficient $\alpha_{V}(T)$ as shown in the inset of Fig.
3. The bulk modulus $B$ is also calculated at different temperatures
according to the formula $$B=V_{0}(\frac{\partial^{2}F}{\partial
V^{2}})|_{V=V_{0}}.$$ For $\alpha$-Pu$_{2}$O$_{3}$, the bulk modulus
decreases with increasing temperature, and the decreasing rate is
much larger than that of PuO$_{2}$, which can be seen from the ratio
of the bulk modulus $B/B_{0}$ as depicted in Fig. 4. Within the
framework of QHA, the considered vibration modes are harmonic but
volume dependent. This volume dependence is somehow neglected in the
above calculations because the considered lattice expansions are
very small.

By using Eqs. (3) and (4), the heat capacity at constant volume
$C_{V}$ can be calculated for $\alpha$-Pu$_{2}$O$_{3}$ and
PuO$_{2}$, which are shown in Fig. 5. Their heat capacity at
constant pressure $C_{p}$ is also calculated from the relationship
$$C_{p}-C_{V}=\alpha_{V}^{2}(T)B(T)V(T)T$$, and shown in Fig. 5. We
can see that as temperature increases, the values of $C_{V}$ and
$C_{p}$ increase continuously with the growth rate of $C_{V}$
slightly less than $C_{p}$, both for $\alpha$-Pu$_{2}$O$_{3}$ and
PuO$_{2}$. Comparing the specific heat of $\alpha$-Pu$_{2}$O$_{3}$
and PuO$_{2}$, there is a crossing point at 265 K. Beneath 265 K,
the increasing rate of heat capacities for $\alpha$-Pu$_{2}$O$_{3}$
is larger than that for PuO$_{2}$. At the room temperature of 300 K,
the values of $C_{V}$ and $C_{p}$ for $\alpha$-Pu$_{2}$O$_{3}$
become 196.1 J kg$^{-1}$ K$^{-1}$ and 199.0 J kg$^{-1}$ K$^{-1}$,
respectively.

\subsection{Thermodynamic equilibrium of different plutonium oxides}

\begin{table*}[ptb]
\caption{Reaction energies $\Delta F$ for the formation of different
plutonium oxides at different temperatures. The parenthesized (Pu)
and (O$_{2}$) represent for the reaction energy for 1 mole Pu atoms
and 1 mole O$_2$ molecules, respectively. For comparison, the
experimental data in Ref. \onlinecite{Haschke} are also listed.}
\label{point}
\begin{ruledtabular}
\begin{tabular}{ccccccccccccc}
&0 K&&300 K&&500 K&&Expt.$^*$\\
\cline{2-9}
Reaction&kcal/mol&kcal/mol&kcal/mol&kcal/mol&kcal/mol&kcal/mol&kcal/mol&kcal/mol\\
&(Pu)&(O$_{2}$)&(Pu)&(O$_{2}$)&(Pu)&(O$_{2}$)&(Pu)&(O$_{2}$)\\
\hline
Pu+O$_{2}$ $\rightarrow$ PuO$_{2}$&-211.0&-211.0&-213.4&-213.4&-215.5&-215.5&-239&-239\\
Pu+$\frac{3}{4}$O$_{2}$ $\rightarrow$
$\frac{1}{2}$Pu$_{2}$O$_{3}$&-176.0&-234.7&-178.0&-237.3&-181.5&-242.0&-189&-252\\
Pu$_{2}$O$_{3}$+$\frac{1}{2}$O$_{2}$ $\rightarrow$ 2PuO$_{2}$&&-139.9&&-137.8&&-136.1&&\\
3PuO$_{2}$+Pu $\rightarrow$ 2Pu$_{2}$O$_{3}$&-71.1&&-74.6&&-79.4&&&\\
\end{tabular}
\begin{flushleft}
$^{*}$ Reference \onlinecite{Haschke}\\
\end{flushleft}
\end{ruledtabular}
\end{table*}

Understanding the thermodynamic equilibrium of different plutonium
oxides at different temperatures can help people get more clear
about the oxidation of plutonium, and further enhance nuclear
security and manipulate the safe maintenance of plutonium.
Chemically, the reaction energy $\Delta F$ can be defined as the
energy differences between initial (reactant) and final (product)
states of a reaction, i.e., $\Delta F =F_{product}-F_{reactant}$. If
a reaction releases energy ($\Delta F$$<$0), it is thermodynamically
favorable and can occur spontaneously. Conversely, if $\Delta F$ is
positive, the reaction is not spontaneous and is hindered until
sufficient energy is added to the system.

By interacting with oxygen molecules, plutonium can be oxidized
according to the processes
\begin{equation}
\mathrm{Pu}+\mathrm{O_{2}} \rightarrow \mathrm{PuO_{2}}, \label{eq5}
\end{equation}
and
\begin{equation}
\mathrm{Pu}+\frac{3}{4}\mathrm{O_{2}} \rightarrow \frac{1}{2}
\mathrm{Pu_{2}O_{3}}. \label{eq6}
\end{equation}
We have known that both $\alpha$-Pu$_{2}$O$_{3}$ and PuO$_{2}$ are
stable in solid phases at room temperature. As the physical and
chemical conditions of the plutonium surface change, these two
oxides are able to convert to each other. When excess O$_{2}$ is
present, $\alpha$-Pu$_{2}$O$_{3}$ is further oxidized through the
process
\begin{equation}
\mathrm{Pu_{2}O_{3}}+\frac{1}{2}\mathrm{O_{2}} \rightarrow
2\mathrm{PuO_{2}}.\label{eq7}
\end{equation}
In contrast, if oxygen is sufficient in the environment, PuO$_{2}$
can be reduced to $\alpha$-Pu$_{2}$O$_{3}$ through the process,
\begin{equation}
3\mathrm{PuO_{2}}+\mathrm{Pu} \rightarrow
2\mathrm{Pu_{2}O_{3}}.\label{eq8}
\end{equation}

Based on their structural similarities, here we only focus on the
transformation between PuO$_2$ and $\alpha$-Pu$_2$O$_3$ without
considering the $\beta$ phase of Pu$_2$O$_3$. During surface
oxidation of plutonium, temperature plays a significant role for the
system to reach thermodynamic equilibrium. For this reason, we
calculated the reaction energies for all the above four processes at
different temperatures, and the results are listed in Table II. As a
comparison, the experimentally determined values by Haschke \emph{et
al.} \cite{Haschke} are also included. In our calculations, bulk
$\delta$-Pu is chosen to be the chemical source of Pu, and its
energy per Pu atoms represents for the chemical potential of Pu. The
chemical potential of oxygen is chosen to be the cohesive energy of
an gas-like O$_2$ molecule. To avoid the overestimation of DFT
methods on the O-O bonding strength, we adopt the experimental
cohesive energy of 5.21 eV \cite{Huber}. From our calculated
results, one can see that all the four reaction processes have
negative reaction energies, indicating that they are
thermodynamically allowable. We can also see from Table II that as
temperature increases, the energy release for reactions (5), (6),
and (8) increases, indicating that these three reactions can be
accelerated by higher temperatures. Especially, we can see that
formation of Pu$_2$O$_3$ is more efficient at higher temperatures.
In contrast, the oxidation reaction of Pu$_2$O$_3$ (reaction (7))
releases more energy at lower temperatures. Thus fully oxidation of
$\delta$-Pu into PuO$_2$ is more efficient at lower temperatures.

The reaction energy values per mole Pu correspond to the situation
where the quantity of Pu is fixed and oxygen is sufficient, while
the values per mole O$_2$ correspond to the situation where the
quantity of oxygen is fixed and Pu is sufficient. We can see from
Table II that the energy release per mole Pu for oxidizing
$\delta$-Pu into PuO$_2$ is always larger than into
$\alpha$-Pu$_2$O$_3$, indicating that formation of PuO$_2$ is more
favorable when oxygen is sufficient in the environment.
Correspondingly, the energy release per mole O$_2$ for oxidizing
$\delta$-Pu into PuO$_2$ is always smaller than into
$\alpha$-Pu$_2$O$_3$, indicating that formation of
$\alpha$-Pu$_2$O$_3$ is more favorable when Pu is sufficient in the
environment. This conclusion accords well with the experimental
results.

\section{Summary}

In summary, we have performed a systematical density functional
theory study to investigate the thermodynamic properties of
$\alpha$-Pu$_{2}$O$_{3}$ and PuO$_{2}$, as well as the thermodynamic
equilibrium between them. The antiferromagnetic order and insulator
character of $\alpha$-Pu$_{2}$O$_{3}$ can be well predicted by the
GGA+$U$ ($U$=3 eV) approach. The theoretical atomic and electronic
structures of the plutonium oxides are in agreement with available
experimental results. Based on the calculated phonon dispersion
curves of $\alpha$-Pu$_{2}$O$_{3}$ and PuO$_{2}$, we systematically
obtain their thermodynamic properties like thermal expansion
coefficient, bulk modulus, and heat capacities at different
temperatures. We find that the volume expansion and bulk modulus of
$\alpha$-Pu$_{2}$O$_{3}$ are both more sensitive to temperature than
PuO$_{2}$, and there is a transition temperature of 265 K comparing
the specific heats of $\alpha$-Pu$_{2}$O$_{3}$ and PuO$_{2}$. In the
temperature range between 0K and 265 K, $\alpha$-Pu$_{2}$O$_{3}$
presents higher heat capacities than PuO$_{2}$. The calculated
reaction energies for the four processes during surface oxidation of
$\delta$-Pu suggest that the reaction products depend heavily on the
temperature and chemical condition of the plutonium surface. The
theoretical reaction energies of Pu oxidizing into
$\alpha$-Pu$_{2}$O$_{3}$ and PuO$_{2}$ are in accord with
experimental data and can reflect the basic qualitative
characteristic of the real physical process. Besides, temperature
plays a significant role in the component of oxidation product. Our
results suggest that when temperature increases, PuO$_{2}$ tends to
be reduced to be $\alpha$-Pu$_{2}$O$_{3}$.

\section{\label{sec:level1}ACKNOWLEDGMENTS}

This work is supported by NSFC under Grant No. 51071032, and by
Foundations for Development of Science and Technology of China
Academy of Engineering Physics under Grants No. 2011A0301016 and No.
2011B0301060.

\end{document}